\documentclass[twocolumn,showpacs,preprintnumbers,amsmath,amssymb]{revtex4}
\begin{document}

\preprint{arXiv: hep-ph/0712.0402} \preprint{CERN-PH-TH/2007-250}

\title{On Naturalness of Scalar Fields and the Standard Model}

\author{Grigorii B. Pivovarov}
\email{gbpivo@ms2.inr.ac.ru}
\affiliation{Institute for Nuclear Research,\\
Moscow, 117312 Russia}

\author{Victor T. Kim}
\email{kim@pnpi.spb.ru}
\affiliation{St. Petersburg Nuclear Physics Institute,\\
Gatchina, 188300 Russia}

\date{December 3, 2007}

\begin{abstract} We discuss how naturalness predicts the scale of
new physics. Two conditions on the scale are considered. The first
is the more conservative condition due to Veltman (Acta Phys.\
Polon.\  B {\bf 12}, 437 (1981)). It requires that radiative
corrections to the electroweak mass scale would be reasonably small.
The second is the condition due to Barbieri and Giudice (Nucl.\
Phys.\  B {\bf 306}, 63 (1988)), which is more popular lately. It
requires that physical mass scale would not be oversensitive to the
values of the input parameters. We show here that the above two
conditions behave differently if higher order corrections are taken
into account. Veltman's condition is robust (insensitive to higher
order corrections), while Barbieri-Giudice condition changes
qualitatively. We conclude that higher order perturbative
corrections take care of the fine tuning problem, and, in this
respect, scalar field is a natural system. We apply the
Barbieri-Giudice condition with higher order corrections taken into
account to the Standard Model, and obtain new restrictions on the
Higgs boson mass. \end{abstract}


\maketitle

It was pointed out in \cite{Susskind:1978ms,'tHooft:1979bh,
Susskind:1982mw} that theories with scalar fields are facing a
serious problem (and the Standard Model is among these). It consists
in absence of a natural explanation for small values of masses of
scalar particles. ("Small" here means much smaller than
the possible fundamental scales like Plank mass or a unification
scale.)

The problem appears as follows. Let us try to expand the physical
mass in a series of bare couplings. In the one-loop approximation we
have \begin{equation} \label{lo} m^2 = m^2_0 + \Lambda^2
P({\lambda_0,g}). \end{equation} Here $m^2$ is the squared mass of a
scalar particle, $m^2_0$ is the corresponding bare mass of the
fundamental Lagrangian of the model defined at the fundamental scale
$\Lambda$, which is also used as a cutoff in the Feynman integrals,
$P({\lambda_0, g})$ is a polynomial of dimensionless bare scalar
field selfcoupling $\lambda_0$ and the rest of dimensionless bare
couplings $g$ of the model, and we neglected the corrections
depending logarithmically on the cutoff. (For example, in the
Standard Model, $P({\lambda_0, g})=
3(3g_2^2+g_1^2+2\lambda_0-4y_t^2)/(32\pi^2)$, where $g_1$, $g_2$,
and $y_t$ are the gauge couplings of the gauge groups $SU(1)$,
$SU(2)$, and top quark Yukawa coupling, respectively
\cite{Veltman:1980mj}.) Here comes the question: How to keep $m$
much less than $\Lambda$? One obvious option is to fine tune the
values of $m^2_0$ and $P({\lambda_0,g})$ to make the two terms in
the right-hand-side of Eq. (\ref{lo}) cancel against each other. But
this seems not to be a natural way (thus the name of the
problem---the naturalness problem). Another way is to ask for a
model where $P({\lambda_0,g})$ is exactly zero (which is the case
for softly broken supersymmetry models \cite{Chung:2003fi}). More
generally, if one rejects unnatural fine tunings of fundamental
parameters, introducing scalar fields one should also point out a
mechanism that keeps the hierarchy between $m$ and $\Lambda$ (the
hierarchy problem).

On a more practical note, Eq. (\ref{lo}) had been used \cite{Veltman:1980mj, Barbieri:1987fn}
to obtain the scale of new physics.
The idea is not to consider $\Lambda$ as a fundamental scale,
but as a scale up to which we can use the low energy effective theory implying Eq. (\ref{lo}).
One may restrict $\Lambda$ requiring, for example \cite{Veltman:1980mj},
that the radiative correction to the mass squared would not exceed the bare mass squared:
\begin{equation}
\label{vc}
|m^2-m^2_0|< m^2_0.
\end{equation}
In what follows we call this condition Veltman's condition.

Another possibility is to restrict not the magnitude of the
radiative correction, but the sensitivity of the physical mass to
small changes in the values of the bare couplings
\cite{Barbieri:1987fn}: \begin{equation} \label{bgc} \Big
|\frac{\lambda_0}{m^2}\frac{\partial m^2}{\partial \lambda_0}\Big |
< q, \end{equation} where $q$ parameterizes the strictness of our
requirements (the value $q = 10$ was suggested in
\cite{Barbieri:1987fn}). Hereafter, we call this condition the
Barbieri-Giudice condition.

Now, assuming that the radiative correction to mass squared is
positive ($P(\lambda_0,g)>0$) and neglecting the difference between
bare and physical couplings, Veltman's condition (\ref{vc}) implies
the following restriction on $\Lambda$: \begin{equation} \label{vcb}
\Lambda^2<\frac{m^2}{2 P(\lambda,g)}, \end{equation} where $\lambda$
denotes the physical coupling corresponding to the bare coupling
$\lambda_0$. The quantities in the right-hand-side of this
inequality are measurable. So we can substitute the measured values,
and obtain an estimate for the scale of new physics. This program
was realized in Ref. \cite{Veltman:1980mj} for the Standard Model.
The outcome is that the scale for the new physics is estimated by
1.2 TeV. Similarly, if we assume Eq. (\ref{lo}), Barbieri-Giudice
condition (\ref{bgc}) implies \begin{equation} \label{bgcb}
\Lambda^2<q \frac{m^2}{\lambda P^\prime(\lambda,g)}, \end{equation}
where the prime over $P$ denotes derivative with respect to
$\lambda$.

As we see, the two conditions yield similar upper bounds for the
scale of new physics. In fact, Veltman's condition  and
Barbieri-Giudice condition are rather different, and the similarity
of the bounds (\ref{vcb}) and (\ref{bgcb}) is due to the use of the
leading order formula (\ref{lo}).

Let us consider what may be the influence of higher order perturbative corrections
on the bounds (\ref{vcb}) and (\ref{bgcb}). This problem was briefly considered in Ref. \cite{Einhorn:1992um}.
It was observed that higher order corrections modify the polynomial $P(\lambda_0,g)$ from (\ref{lo})
(even making it dependent on $\Lambda$ logarithmically). If this would be the only way higher order corrections
are getting involved, they could not influence significantly the bounds (\ref{vcb}) and (\ref{bgcb})
(at least, at small couplings).

Unfortunately, there are important higher order corrections overlooked in Ref. \cite{Einhorn:1992um}:
In higher orders of the expansion of the physical mass squared in powers of the bare couplings, Eq. (\ref{lo}),
higher powers of $\Lambda$ will appear, and the larger the order of perturbation theory,
the larger is the power of $\Lambda$ appearing in the right-hand-side of Eq. (\ref{lo}).
For example, in the third order in $\lambda_0$ there is a diagram with two tadpoles attached
to the scalar propagator. It gives contribution proportional to $\lambda_0^3\Lambda^4/m_0^2$.
Similarly, in the expansion of the physical couplings in powers of bare couplings,
infinitely high powers of $\Lambda$ appear, and the power of $\Lambda$ appearing
in the expansion is bounded only if we consider a finite order of the perturbation theory in $\lambda_0$.

A direct approach is to study the powers of $\Lambda$ appearing in the expansion of
physical parameters in powers of bare couplings. This may be an interesting problem,
but there is a shortcut allowing one to avoid it. Indeed, for renormalizable theories,
dependence of {\it bare couplings} on the cutoff is known if they are expressed
in terms of the {\it physical couplings} \cite{Bogolyubov:1980nc}.
Let us reiterate: for renormalizable theory, {\it bare} mass squared of a scalar particle
expressed as a series in powers of {\it physical} couplings with coefficients of
the expansion depending on the cutoff, {\it physical} masses and renormalization scale
grows not faster than the cutoff squared. Is this statement compatible with the appearance
of higher powers of the cutoff in the right-hand-side of Eq. (\ref{lo})? It is easy to check
that there is no contradiction. Indeed, schematically, if we take the renormalization scale
to be of the order of physical mass, the bare mass squared and the bare coupling are expressed as follows
\begin{eqnarray}
\label{mr}
m^2_0 &=& m^2 - \Lambda^2\, P(\lambda, g),    \\
\label{cr} \lambda_0\, &=&\lambda + \log(\frac{\Lambda^2}{m^2})\,
\frac{\beta(\lambda,g)}{2}, \end{eqnarray} where $P(\lambda,g)$ is
(in the leading order) the same polynomial as in Eq.(\ref{lo}), and
$\beta(\lambda,g)$ is the leading order of the beta function
governing the renormalization group evolution of coupling $\lambda$.
If we use the above expressions as equations for $m^2$ and
$\lambda$, we can determine the expansions of $m^2$ and $\lambda$ in
powers of $\lambda_0$. It is easy to check that both power series
involve arbitrary high powers of the cutoff. The reason for the
appearance of the high powers of $\Lambda$ in the expansions is the
presence of $m^2$ in the argument of the logarithm. (Logarithmic
term is also present in the formula for bare mass, but we dropped
it, because it is insignificant for further reasoning.)

If we put $\beta(\lambda,g)=0$ in Eq. (\ref{cr}), we derive the
bounds (\ref{vcb}) and (\ref{bgcb}) from Veltman's condition
(\ref{vc}) and Barbieri-Giudice condition (\ref{bgc}), respectively.
Evidently, the bound (\ref{vcb}) is not influenced by nonzero
$\beta(\lambda,g)$ in any way. In what follows, we see how the fact
that $\beta(\lambda,g)\neq 0$ influences the bound (\ref{bgcb}).

We need to compute the derivative $\partial m^2/\partial \lambda_0$
involved in Barbieri-Giudice condition (\ref{bgc}). More generally,
we need to compute the entries of the matrix \begin{equation}
\label{matra} A = \left( \begin{array}{cc}
\frac{\partial \lambda}{\partial \lambda_0}&\frac{\partial \lambda}{\partial m_0^2}\\
\frac{\partial m^2}{\partial \lambda_0}&\frac{\partial m^2}{\partial
m_0^2} \end{array}\right). \end{equation} The inverse of the desired
$A$ can be computed with Eqs. (\ref{mr}) and (\ref{cr}):
\begin{eqnarray} \label{matrb} A^{-1}\equiv B &=& \left(
\begin{array}{cc}
\frac{\partial \lambda_0}{\partial \lambda}&\frac{\partial \lambda_0}{\partial m^2} \\
\frac{\partial m_0^2}{\partial \lambda}&\frac{\partial m_0^2}{\partial m^2} \end{array}\right)\\
&=&\left( \begin{array}{cc}
1+\log(\frac{\Lambda^2}{m^2})\frac{\beta^\prime(\lambda,g)}{2}&-\frac{\beta(\lambda,g)}{2m^2}\\
-\Lambda^2 P^\prime(\lambda,g)&1 \end{array}\right), \end{eqnarray}
where primes over $\beta$ and $P$ denote the derivative with respect
to $\lambda$. Thus, the desired $A$ is \begin{equation}
\label{matrae} A = \frac{1}{\det(B)}\left( \begin{array}{cc}
1&\frac{\beta(\lambda,g)}{2m^2}\\
\Lambda^2 P^\prime(\lambda,g)&
1+\log(\frac{\Lambda^2}{m^2})\frac{\beta^\prime(\lambda,g)}{2}\end{array}\right),
\end{equation} where \begin{equation} \label{det} \det(B) =
-\frac{\Lambda^2}{m^2}P^\prime(\lambda,g)\frac{\beta(\lambda,g)}{2}
+ \log(\frac{\Lambda^2}{m^2})\frac{\beta^\prime(\lambda,g)}{2} + 1.
\end{equation} Now we see why it is important to keep nonzero
$\beta(\lambda,g)$ in the consideration: Neglecting
$\beta(\lambda,g)$ removes the most important first two terms in the
right-hand-side of this expression. As a consequence, neglecting
$\beta(\lambda,g)$ leads to a qualitative mistake in the estimate of
the behavior of the matrix of derivatives $A$ in the limit of large
$\Lambda$.

Finally, in the limit of infinite $\Lambda$, we have:
\begin{equation}
\label{matrall}
A = \left( \begin{array}{cc}
0&0\\
-\frac{2m^2}{\beta(\lambda,g)}& 0\end{array}\right).
\end{equation}

Let us comment on Eq. (\ref{matrall}). As we see, physical
parameters---the observable mass and coupling---are not
oversensitive to the values of the bare parameters defined at a
large (e.g., fundamental) scale $\Lambda$. The leading order
relation, Eq. (\ref{lo}), is misleading in this respect. In other words:
 Derivative of observable mass in bare
coupling has a finite limit expressible in terms of observable
parameters when the cutoff is removed. ( This is the worst
sensitivity we have: the physical coupling exhibits {\it
universality}, i.e., it becomes independent of bare parameters at
infinite cutoff; the physical mass becomes independent of the bare
mass at infinite cutoff.) We conclude that the fine tuning
problem is the problem of the leading order perturbative
approximation, Eq. (\ref{lo}).

Now we can derive from the Barbieri-Giudice condition(\ref{bgc}) the
inequality \begin{equation} \label{bgcbc} \big |
\frac{2\lambda}{\beta(\lambda,g)} \big | < q, \end{equation} where
we neglected the difference between $\lambda$ and $\lambda_0$.

Let us specialize inequality (\ref{bgcbc}) to the case of the
Standard Model. The Standard Model one-loop beta-function governing
the evolution of the scalar selfcoupling $\lambda$ is
\cite{Machacek:1981ic} \begin{eqnarray}
\label{betasm}
\beta(\lambda,g) & = & \frac{6}{8\pi^2}(\lambda^2-\lambda
[\frac{1}{4}g_1^2+\frac{3}{4}g_2^2-g_t^2]
\nonumber\\
&+&\frac{1}{16}g_1^4
+\frac{1}{8}g_1^2g_2^2+\frac{3}{16}g_2^4-y_t^4), \end{eqnarray}
where $g_1$ and $g_2$ are gauge couplings of $SU(1)$ and $SU(2)$
respectively, $y_t = m_t/v$ ($m_t$ is the mass of the top quark, and
$v$ is the vacuum expectation of the scalar field). The couplings
involved in the expression for the beta function can be expressed
via ratios of the masses and the scalar field vacuum expectation
value $v$. In this way, for the Standard Model, Barbieri-Giudice
condition (\ref{bgc}) implies the following inequality:
\begin{equation} \label{smbgc}
\frac{4m_H^2v^2}{|p(m_H,m_Z,m_W,m_t)|}<\frac{3q}{4\pi^2},
\end{equation} where $p(m_H,m_Z,m_W,m_t)$ is the following
polynomial of the Higgs, $Z$, $W$ and top quark masses:
\begin{eqnarray} \label{p} p(m_H,m_Z,m_W,m_t) & = & m_H^4 +
m_H^2(2m_t^2-m_Z^2-2m_W^2)  \nonumber\\
& -& 4m_t^4 + m_Z^4+2m_W^4. \end{eqnarray} Thus, Barbieri-Giudice
condition (\ref{bgc})  implies a restriction on the Higgs boson
mass. Using known values, we see that inequality (\ref{smbgc})
forbids moderate values of the Higgs boson mass. For example, if we
take $q=10$, we obtain that the band of values of $m_H$
approximately from 96 GeV to 540 GeV is forbidden. (The value for
the upper boundary of the forbidden band is hardly reliable, because
it corresponds to strongly interacting Higgs bosons.) If we relax
the Barbieri-Giudice condition and take $q = 15$ (20), the forbidden
band shrinks: it ranges from 113 (126) GeV to 438 (380) GeV.

Let us summarize our findings. Taking into account
higher order perturbative corrections does not change the basic
fact: radiative corrections to the electroweak scale are growing
fast with cutoff. At 1.2 TeV the correction to the intermediate
bosons mass squared is about a half of the total mass squared. Is it
new physics that half of the observable mass scale is due to
radiative corrections is a matter of convention. We consider such a
situation as deserving the title of new physics. To say the least,
perturbation theory looks jeopardized in such
circumstances. Beyond perturbation theory, we still do
not know any mechanism that would provide for small masses of the
scalar particles.

On the other hand, if some unknown mechanism provides for small mass of scalar particles, perturbation theory is quite able to  explain relative stability of the scalar mass against small variations in fundamental parameters. We demonstrated that there is no fine tuning problem
in the theory of quantum scalar field,
and derived inequality (\ref{smbgc}) in the Standard Model
restricting the Higgs boson mass. Phenomenological consequences
of this restriction will be studied elsewhere.

We thank CERN Theory Unit, where a part of the research
was performed, for its warm hospitality.


\begin{thebibliography}{11}
\expandafter\ifx\csname natexlab\endcsname\relax\def\natexlab#1{#1}\fi
\expandafter\ifx\csname bibnamefont\endcsname\relax
  \def\bibnamefont#1{#1}\fi
\expandafter\ifx\csname bibfnamefont\endcsname\relax
  \def\bibfnamefont#1{#1}\fi
\expandafter\ifx\csname citenamefont\endcsname\relax
  \def\citenamefont#1{#1}\fi
\expandafter\ifx\csname url\endcsname\relax
  \def\url#1{\texttt{#1}}\fi
\expandafter\ifx\csname urlprefix\endcsname\relax\def\urlprefix{URL }\fi
\providecommand{\bibinfo}[2]{#2}
\providecommand{\eprint}[2][]{\url{#2}}

\bibitem{Susskind:1978ms}
  K.~G.~Wilson,
  Phys.\ Rev.\  D {\bf 3}, 1818 (1971); \\
  L.~Susskind,
  Phys.\ Rev.\  D {\bf 20}, 2619 (1979).

\bibitem{'tHooft:1979bh}
  G.~'t Hooft,
  Proc. Cargese Summer Inst., Gargese, France, Aug. 26 - Sep. 8, \ 1979,
  ed: G.~'t Hooft et al. (NY, Plenum Press, 1980) 135.

\bibitem{Susskind:1982mw}
  L.~Susskind,
  Phys.\ Rept.\  {\bf 104}, 181 (1984).

\bibitem{Chung:2003fi}
  D.~J.~H.~Chung, L.~L.~Everett, G.~L.~Kane, S.~F.~King, J.~D.~Lykken and L.~T.~Wang,
  Phys.\ Rept.\  {\bf 407}, 1 (2005).


\bibitem{Veltman:1980mj}
  M.~J.~G.~Veltman,
  Acta Phys.\ Polon.\  B {\bf 12}, 437 (1981).

\bibitem{Barbieri:1987fn}
  R.~Barbieri and G.~F.~Giudice,
  Nucl.\ Phys.\  B {\bf 306}, 63 (1988).

\bibitem{Bogolyubov:1980nc}
  N.~N.~Bogolyubov and D.~V.~Shirkov,
  Intersci.\ Monogr.\ Phys.\ Astron.\  {\bf 3}, 1 (1959).



\bibitem{Einhorn:1992um}
  M.~B.~Einhorn and D.~R.~T.~Jones,
  Phys.\ Rev.\  D {\bf 46}, 5206 (1992).

\bibitem{Machacek:1981ic}
  M.~E.~Machacek and M.~T.~Vaughn,
  Phys.\ Lett.\  B {\bf 103}, 427 (1981).





\end{thebibliography}
\end{document}